\documentclass[twocolumn]{aastex701}
\usepackage{amsmath}

\begin{document}

\title{Detection of an Extended Ly$\alpha$ Halo around a $\textit{z}=6.64$ Broad Absorption Line Quasar with the Keck Cosmic Web Imager}

\correspondingauthor{Raymond~P.~Remigio}

\author[0000-0002-0164-8795,gname='Raymond', sname='Remigio']{Raymond~P.~Remigio}
\affiliation{Department of Physics and Astronomy, 4129 Frederick Reines Hall, University of California, Irvine, CA 92697, USA}
\email[show]{remigior@uci.edu}

\author[0000-0002-3026-0562,gname='Aaron',sname='Barth']{Aaron~J.~Barth}
\affiliation{Department of Physics and Astronomy, 4129 Frederick Reines Hall, University of California, Irvine, CA 92697, USA}
\email{barth@uci.edu}

\author[0000-0002-7633-431X,gname='Feige',sname='Wang']{Feige~Wang}
\affiliation{Department of Astronomy, University of Michigan, 1085 S. University Ave., Ann Arbor, MI 48109, USA}
\email{fgwang.astro@gmail.com}

\author[0000-0001-5287-4242,gname='Jinyi',sname='Yang']{Jinyi~Yang}
\affiliation{Department of Astronomy, University of Michigan, 1085 S. University Ave., Ann Arbor, MI 48109, USA}
\email{jyyangas@umich.edu}

\author[0000-0002-7054-4332,gname='Joseph', sname='Hennawi']{Joseph~F.~Hennawi}
\affiliation{Leiden Observatory, Leiden University, P.O. Box 9513, 2300 RA Leiden, The Netherlands}
\affiliation{Department of Physics, University of California, Santa Barbara, CA 93106, USA}
\email{joe@physics.ucsb.edu}

\author[0000-0001-7653-5827,gname='Ryan',sname='Cooke']{Ryan~J.~Cooke}
\affiliation{Centre for Extragalactic Astronomy, Department of Physics, Durham University, South Road, Durham DH1 3LE, UK}
\email{ryan.j.cooke@durham.ac.uk}

\author[0000-0002-2931-7824, gname='Eduardo', sname='Ba\~{n}ados']{Eduardo~Ba\~{n}ados}
\affiliation{Max-Planck-Institut f\"ur Astronomie, K\"onigstuhl 17, D-69117 Heidelberg, Germany}
\email{banados@mpia.de}

\author[0000-0003-3310-0131,gname='Xiaohui', sname='Fan']{Xiaohui~Fan}
\affiliation{Steward Observatory, University of Arizona, 933 North Cherry Avenue, Tucson, AZ 85721, USA}
\email{xfan@arizona.edu}

\author[0000-0002-6822-2254, gname='Emanuele', sname='Farina']{Emanuele~Paolo~Farina}
\affiliation{International Gemini Observatory/NSF NOIRLab, 670 N. A’ohoku Place, Hilo, HI 96720, USA}
\affiliation{INAF -- Osservatorio di Astrofisica e Scienza dello Spazio di Bologna via Gobetti 93/3, I-40129, Bologna, Italy}
\email{emanuele.farina@noirlab.edu}

\begin{abstract}
We present the first results from a program searching for extended Ly$\alpha$ halos around high redshift ($ z \gtrsim 6.5$) quasars using the red channel of the Keck Cosmic Web Imager (KCWI).
Our observations reveal a Ly$\alpha$ halo extending to $\simeq11$ pkpc around the $z=6.64$ broad absorption line quasar J0910$-$0414.
The Ly$\alpha$ velocity field displays a rotation-like gradient, and the gas velocity dispersion is consistent with gravitationally dominated motion ($\sigma_{\mathrm{Ly\alpha}}<300$ km s$^{-1}$).
Comparison with the [\ion{C}{2}] kinematics of the host galaxy core from ALMA observations shows that the Ly$\alpha$-emitting gas extends over a much larger region, shows distinct kinematics, and has a smaller velocity dispersion ($\sigma_{\mathrm{Ly\alpha}} \simeq 0.6\sigma_{\mathrm{[C\;II]}}$).
The Ly$\alpha$ spectral region of the quasar is largely obscured by a deep \ion{N}{5} absorption trough, and as a result, roughly 55\% of the total Ly$\alpha$ flux is from the extended halo.
These observations demonstrate the potential of KCWI for probing the cool gas reservoir that fuels the growth of quasars and their hosts in the epoch of reionization.

\end{abstract}



\section{Introduction}

The discovery of quasars beyond $z > 6$ \citep{fan01} has presented challenges to theories of early supermassive black hole (SMBH) formation and growth. 
Recent observations of these reionization-era quasars have established the presence of SMBHs with estimated masses on the order of $10^{9}$ $\mathrm{M}_\odot$ \citep[e.g.,][]{shen19, onoue19, yang21, farina22, belladitta25} within a billion years of the Big Bang.
Their discovery raises questions about how early SMBHs can grow to billions of solar masses in such a short time, with current theories requiring black hole seeds to undergo periods of sustained accretion at the Eddington rate (for heavy seeds), or short bursts of super-Eddington accretion \citep[e.g.,][]{volonteri06, fan23,alexander25}. 
Through mm and sub-mm observations, these quasar hosts have been found to display high dust masses and rapid star formation \citep[e.g.,][]{venemans16, decarli18, wang24b}.
Measurements of SMBH mass and star formation rate imply that these quasars must reside in environments with an ample supply of cool gas to fuel star formation and SMBH growth. 
Hydrodynamical simulations show that the most massive black holes at $z\sim6$ ($\sim10^{9}$ $\mathrm{M}_\odot$) reside in dark matter halos at the intersection of gas filaments within overdense regions in the early Universe \citep{costa14}, and that radiation from the quasar can interact with the cool gas, leading to the appearance of an extended Ly$\alpha$ halo \citep[e.g.,][]{haiman01, costa22, fan23}. 
Targeted studies of extended Ly$\alpha$ emission can provide important insights about the kinematics and spatial distribution of the cool gas supply, as well as the orientation of the quasar itself \citep{den_brok20, zhang25}. 

\added{Extended Ly$\alpha$ nebulae around quasars have been studied through narrow-band imaging \citep[e.g.,][]{cantalupo14,hennawi15,momose19} and long-slit spectroscopic \citep[e.g.,][]{roche14, hoshi25} observations, with some nebulae displaying spatial extents exceeding 100 pkpc.
The development of integral-field spectrographs on 8-10 m class telescopes has enabled simultaneous spatial and spectroscopic information to be gathered within a single set of observations, allowing for more complete and detailed characterization of the extended emission.} 
Ground-based surveys utilizing integral-field spectrographs have searched for extended Ly$\alpha$ halos around quasars at $z\sim2$\textendash$6$ \citep[e.g.,][]{borisova16, farina17, arrigoni19, drake19, cai19, farina19, herwig24, lobos25}, with detections of up to $z\sim2$ achieved with the Keck Cosmic Web Imager \citep[KCWI;][]{morrissey18}, and up to $z=6.62$ \citep{farina17, farina19} at the VLT with the Multi Unit Spectroscopic Explorer \citep[MUSE;][]{bacon10}.
To date, MUSE has been at the forefront of searches for extended Ly$\alpha$ around distant quasars due to its extensive wavelength coverage, but its cutoff at 9300 \AA~limits searches to $z\lesssim6.6$.
\added{In order to trace the physical processes responsible for early SMBH growth and evolution, it is necessary to push Ly$\alpha$ searches to higher redshifts and probe the first generation of quasars \citep{wang19}.}

The red arm of KCWI, also known as the Keck Cosmic Reionization Mapper \citep[KCRM;][]{mcgurk24, kassis2024}, was commissioned in semester 2023A, with routine observations starting in 2023B. 
\added{The red channel extends KCWI's effective wavelength coverage out to $\approx 1$ $\mu$m, and its sensitivity beyond 9000 \AA~enables searches for Ly$\alpha$ halos out to $z = 7.1$ at depths comparable to the deepest MUSE observations at $z\sim6$.}
We initiated a program in semester 2023B to search for extended Ly$\alpha$ around $z > 6.5$ quasars, including sources selected from the sample of the JWST ASPIRE program \citep[GO-2078, PI: Feige Wang;][]{wang23}, which obtained slitless spectroscopy covering the H$\beta$+[\ion{O}{3}] spectral region at the quasars' redshifts.

In this Letter, we present KCWI observations of J0910$-$0414, a $z=6.64$ ASPIRE quasar.
J0910$-$0414 is a broad absorption line (BAL) quasar discovered by \cite{wang19}, 
characterized by its deep absorption troughs blueward of \ion{N}{5} and \ion{C}{4} \citep{yang21}.
ALMA [\ion{C}{2}] observations yield a systemic redshift of $z_{[\mathrm{C\;II}]} = 6.6363 \pm 0.0003$ \citep{wang24b}. 
Near-infrared (NIR) observations presented by \cite{yang21} yield a \ion{Mg}{2}-based black hole mass of $M_{\mathrm{BH}} = (3.59 \pm 0.61) \times 10^{9}\;\mathrm{M_\odot}$, one of the most massive SMBHs known at $z>6.5$ \citep{wang24a}.
Sub-mm observations also suggest that J0910$-$0414 hosts extreme star formation activity, with star formation rate (SFR) estimates ranging from $\sim 100$ to $\sim 1000 \; \mathrm{M}_{\odot}\; \mathrm{yr}^{-1}$ depending on the tracer \citep{wang24b}.
Recent studies of the quasar environment further reveal that J0910$-$0414 resides at the center of a dense protocluster, with multi-wavelength observations confirming the presence of 12 Ly$\alpha$ and three [\ion{C}{2}] emitters, one of which is interacting with the quasar at a projected separation of $\simeq3$ proper kiloparsecs \citep[pkpc;][]{wang24a}.

This paper is organized as follows. We describe the observations and data reduction in Section \ref{sec:red}.
In Section \ref{sec:detect}, we detail our methods used to subtract the quasar emission and isolate the extended Ly$\alpha$, and create maps of the spatially resolved kinematics. 
We interpret our results in Section \ref{sec:disc}, and conclude the paper in Section \ref{sec:end}. 
Throughout this paper, we adopt a flat $\Lambda$CDM cosmology with $H_0 = 70$ km s$^{-1}$ Mpc$^{-1}$, $\Omega_{\mathrm{m}} = 0.3$, and $\Omega_{\Lambda} = 0.7$. 
At $z=6.64$, the Universe is 0.807 Gyr old, and an angular distance of $1\arcsec$ corresponds to $5.4$ pkpc \citep[][]{wright2006}. 

\begin{figure*}[t]
    \centering
    \includegraphics[width=1\textwidth]{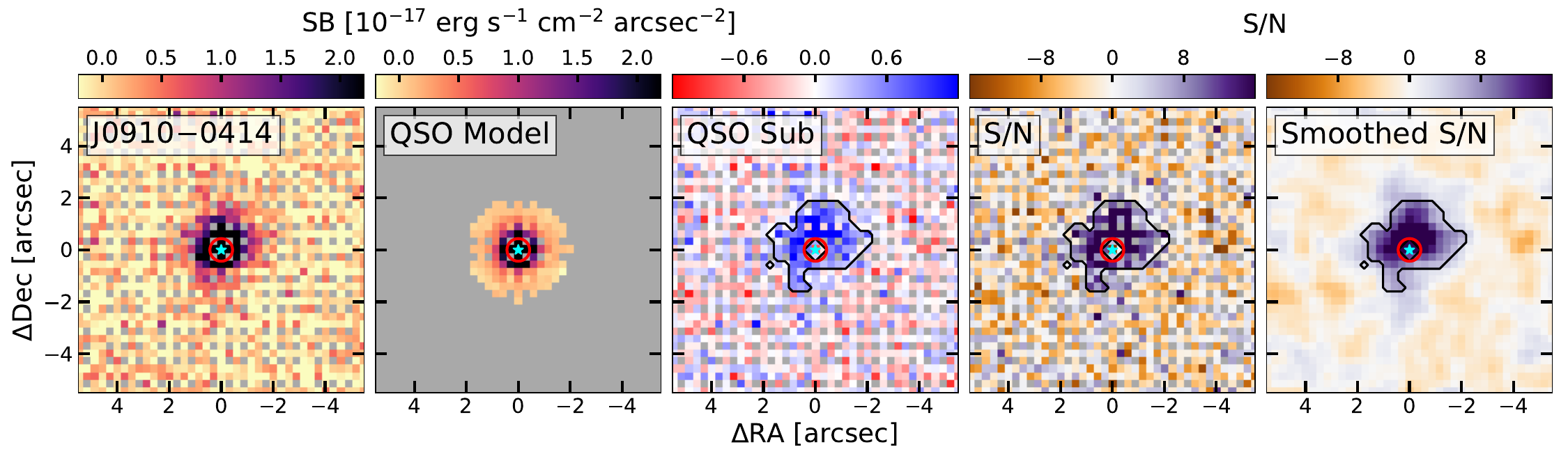}
    \caption{Data, QSO point source model, QSO-subtracted, S/N, and smoothed S/N cubes of J0910$-$0414, as described in Section \ref{sec:detect}. The images in each panel were formed by collapsing each cube over the spectral range where extended Ly$\alpha$ is detected (i.e., from $\lambda_{\mathrm{min}}$ to $\lambda_{\mathrm{max}}$). Gray pixels mark spaxels where no flux from the detector pixels was assigned when creating the cubes. The cyan star denotes location of the quasar, determined from the peak emission in a continuum white light image, and the red circle encloses the pixels that were used to scale the PSF model. The black contours in the QSO-subtracted, S/N, and smoothed S/N plots outline the 2D spatial mask described  in Section \ref{sec:detect}.
    Surface brightness values for the three leftmost panels have not been corrected for cosmological dimming or extinction.
    In all panels, North is up and East is toward the left.}
    \label{fig:j0910_multi_panel}
\end{figure*}

\section{Observations and Data Reduction} \label{sec:red}

We observed J0910$-$0414 with KCWI on the Keck II telescope as part of Program 2024A\_N061 (PI: Feige Wang). 
Our instrument configuration uses the medium slicer with the RM2 grating and $2\times2$ binning, which yields a field of view (FoV) of 16\farcs{5} $\times$ 20\farcs{4}, a wavelength coverage of 
8500\textendash10500 \AA, a slice width of $0\farcs{679}$, and a detector plate scale of $0\farcs{291}$ px$^{-1}$. 
We chose a central wavelength of 9500 \AA~so that the Ly$\alpha$ emission line at a redshift of $z=6.6$ lies near the center of the observed wavelength coverage.
From FeAr arc line exposures, we measured an average full-width at half-maximum (FWHM) of 2.4 \AA~across all slices, which translates to an instrumental dispersion of $\sigma_{\mathrm{inst}}\approx 32$ km s$^{-1}$.

We observed J0910$-$0414 at five dither positions (initial pointing plus pointings on the corners of a 1\farcs{0} $\times$ 1\farcs{0} square centered on the quasar), with this dither pattern executed at position angles (PAs) of $0^{\circ}$ and $90^{\circ}$, to ensure equal spatial sampling in the horizontal and vertical directions for the final data cube.
For each dither position and PA, we took a series of four 300 s exposures to ensure better cosmic ray rejection.
Altogether, our data constitutes $3.67$ h of integration time, with an observed $5\sigma$ surface brightness limit of $2 \times 10^{-18}$ erg s$^{-1}$ cm$^{-2}$ arcsec$^{-2}$ in a 200 km s$^{-1}$ spectral bin at the expected wavelength of Ly$\alpha$, calculated over an area of 1 arcsec$^{2}$.
From white light cubes created for individual exposure sets, we measure seeing values between 0\farcs{9} and 1\farcs{3}, with a mean value of 1\farcs{0}.

We reduced the data using \texttt{PypeIt}\footnote{ \url{https://github.com/pypeit/PypeIt} }\citep[][R. J. Cooke et al. in prep.]{pypeit_release, pypeit_joss}, which performs bias subtraction, flat-fielding, wavelength calibration, flux calibration, and data cube creation.
\texttt{PypeIt} converts the reduced 2D spectra to data cubes using the nearest grid point (NGP) algorithm, which assigns all of the flux from a detector pixel into a single voxel (coordinates of $\lambda$, RA, Dec) in the output data cube.
As a result, noise in the data cubes is not correlated between voxels.
We used this method to create cubes and continuum white light images for each exposure set.
The continuum white light images were used to refine the relative spatial offsets between the different cubes, which were then passed to \texttt{PypeIt} to produce the final data cube. 
The data cube covers a 17\farcs{0} $\times$ 17\farcs{0} FoV centered on the quasar, with 0\farcs{291} $\times$ 0\farcs{291} $\times$ 0.966 \AA~voxels.
We compared the RMS spectrum in an empty sky region of the coadded cube to the mean spectrum in the same spatial region of the 1$\sigma$ error cube, and calculated a wavelength-dependent correction factor for the noise, which was then applied to the error cube.
Lastly, we applied a telluric correction using \texttt{PypeIt}'s \texttt{tellfit} routine.

\begin{figure*}[t]
    \centering
    \includegraphics[width=1\textwidth]{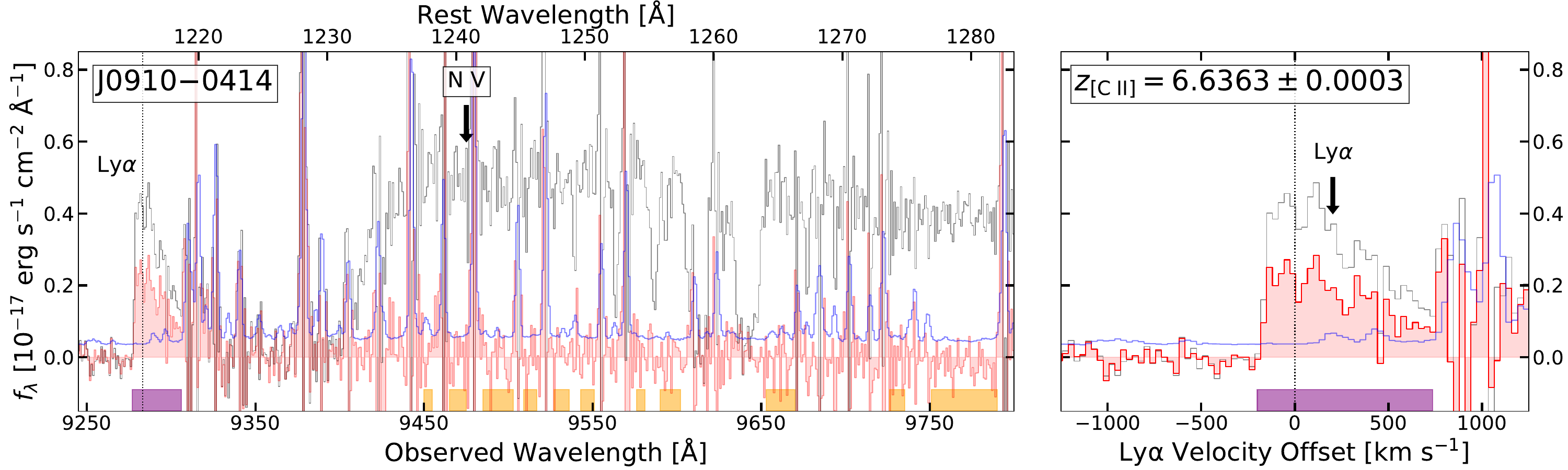}
    \caption{
    Left panel: The total flux (gray), extended Ly$\alpha$ (red), and \added{$1\sigma$ error} (blue) extracted over the 2D spatial mask as described in Section \ref{sub:halo_spec}.
    The vertical dotted line denotes the expected wavelength of Ly$\alpha$ at the quasar's [\added{\ion{C}{2}}] redshift. 
   The black arrow marks the location of \ion{N}{5}.
    The purple band displays the spectral range $(\lambda_{\mathrm{min}},\lambda_{\mathrm{max}})$ over which extended Ly$\alpha$ was detected. The orange bands show the spectral ranges used to create the PSF model.
   Right panel: Total flux density, extended Ly$\alpha$, and \added{$1\sigma$ error} as a function of the velocity offset relative to Ly$\alpha$ at the quasar's [\added{\ion{C}{2}}] redshift. 
   The black arrow indicates the Ly$\alpha$ velocity centroid of the halo.}
    \label{fig:neb_spec}
\end{figure*}

\section{Isolating the Extended Emission} \label{sec:detect}
Detection of the faint, extended Ly$\alpha$ emission requires subtraction of the bright, spatially unresolved emission from the quasar. 
We performed a procedure to search for the extended emission similar to the method used by the Reionization Epoch QUasar InvEstigation with MUSE \citep[REQUIEM;][]{farina19} survey.

We created a model of the point-spread function (PSF) by first collapsing the cube over the spectral region $9450$\textendash $10200$ \AA, excluding any wavelength channels containing strong absorption features or sky subtraction artifacts.
We then fit a 2D Gaussian to the quasar in the resulting white light image and cropped the image to a circular stamp with a radius equal to five times the Gaussian $\sigma$.
The stamp serves as an empirical model of the PSF, which is assumed to be free of extended emission and contains negligible contribution from the host galaxy.
For each wavelength channel, we scaled the PSF model according to the total flux within a 1-pixel radius of the quasar center, which we assume to be dominated by the unresolved AGN emission. 
We created a cube of the quasar model and subtracted this from our primary data cube, resulting in a cube that contains only extended emission. The collapsed science, PSF model, and PSF-subtracted cubes are displayed in Figure \ref{fig:j0910_multi_panel}.

To search for extended Ly$\alpha$ in our PSF-subtracted cube, we follow \cite{hennawi_prochaska_13} and create a separate cube of the smoothed signal-to-noise (S/N), using a 3D Gaussian kernel with $\sigma_{\mathrm{spat}} = 0\farcs{291}$ and $\sigma_{\mathrm{spec}}=0.966$ \AA, equivalent to the dimensions of a single voxel in the data cube.
We apply a friend-of-friends (FoF) clustering algorithm on the smoothed S/N cube to identify the voxels with significant extended emission.
First, we identify the most significant voxel in the smoothed cube within a 500 km s$^{-1}$ window of the \added{expected wavelength of Ly$\alpha$ at the quasar's [\added{\ion{C}{2}}] redshift (chosen to encompass the typical Ly$\alpha$ halo centroid velocity offsets measured in the REQUIEM survey)}, and within $1\arcsec$ of the quasar center. 
Then, we subsequently searched for other nearby voxels, requiring that they are within a distance of two voxels (in pixel coordinates) of another significant voxel, and are above a certain S/N threshold $\eta$, this time defined to be the ratio between the value of the smoothed cube in a single voxel and the standard deviation of the smoothed S/N cube over an empty spectral range blueward of Ly$\alpha$.
We perform this process for significance thresholds of $\eta = \{2.0, 3.0, 4.0, 5.0\}$, creating four separate 3D masks identifying pixels with significant extended emission. 

To determine the optimal significance threshold, we used the 3D masks to create 2D spatial masks identifying spaxels containing significant Ly$\alpha$ emission. We imposed the additional criterion that the spaxel must have at least two consecutive significant voxels in the spectral direction in the 3D mask. 
We then updated the 3D mask to exclude any voxels located in spaxels that were excluded from the 2D spatial mask.
We used the 2D masks to extract spectra of the extended Ly$\alpha$ emission and calculate the S/N of the line flux, integrating over a wavelength range determined by the minimum and maximum wavelengths of significant voxels in the 3D mask. 
Among our significance thresholds, $\eta = 3.0$ yielded the maximum S/N of the integrated line flux with $\mathrm{S/N} = 17.0$, and we adopt the 2D spatial mask and wavelength limits ($\lambda_{\mathrm{min}}=9277.1$, $\lambda_{\mathrm{max}}=9306.1$) for this threshold for the remainder of the paper. 
The S/N cube, its smoothed version, and the 2D spatial mask are displayed in Figure \ref{fig:j0910_multi_panel}. 
We measured three characteristic sizes of the nebula from the 2D mask:
the largest separation between two spaxels within the mask $d_{\mathrm{max}}$,
the distance from the center of the quasar to the farthest spaxel within the mask $d_{\mathrm{r}}$,
and the area of the extended emission $A_{\mathrm{Ly\alpha}}$. 
These measurements are listed in Table \ref{tab:halo}.

\subsection{Spectrum of the Ly$\alpha$ Halo}
\label{sub:halo_spec}

We 
\added{corrected our cubes for Galactic extinction} 
with the \texttt{dust\_extinction}\footnote{\url{https://dust-extinction.readthedocs.io/en/latest/}} \citep{Gordon2024} package, using $E(B-V) = 0.0225$ \citep{schlafly_finkbeiner_11} and the \cite{gordon23} extinction curve with $R_{\mathrm{V}} = 3.1$.
Next, we used the 2D spatial mask described in Section \ref{sec:detect} to extract spectra of the total flux and extended halo.
We calculate the Ly$\alpha$ integrated line flux, velocity and dispersion as the zeroth, first, and second velocity moments, respectively \citep[e.g., Equations 1\textendash5 of][]{remigio25}.
Additionally, we estimate the FWHM of Ly$\alpha$ in the halo spectrum after smoothing with a 1D Gaussian kernel with $\sigma_{\mathrm{spec}} = 0.966$ \AA.
We calculate the velocity moments non-parametrically, by direct integration from $\lambda_{\mathrm{min}}$ to $\lambda_{\mathrm{max}}$. 
We correct for the instrumental dispersion by subtracting the measured KCWI spectral resolution in quadrature.
The spectra are displayed in Figure \ref{fig:neb_spec}, and measurements of the extended emission are included in Table \ref{tab:halo}.

\subsection{The Ly$\alpha$ Radial Surface Brightness Profile}
\label{sub:radial profile}
\begin{figure}
    \centering
    \includegraphics[width=0.479\textwidth]{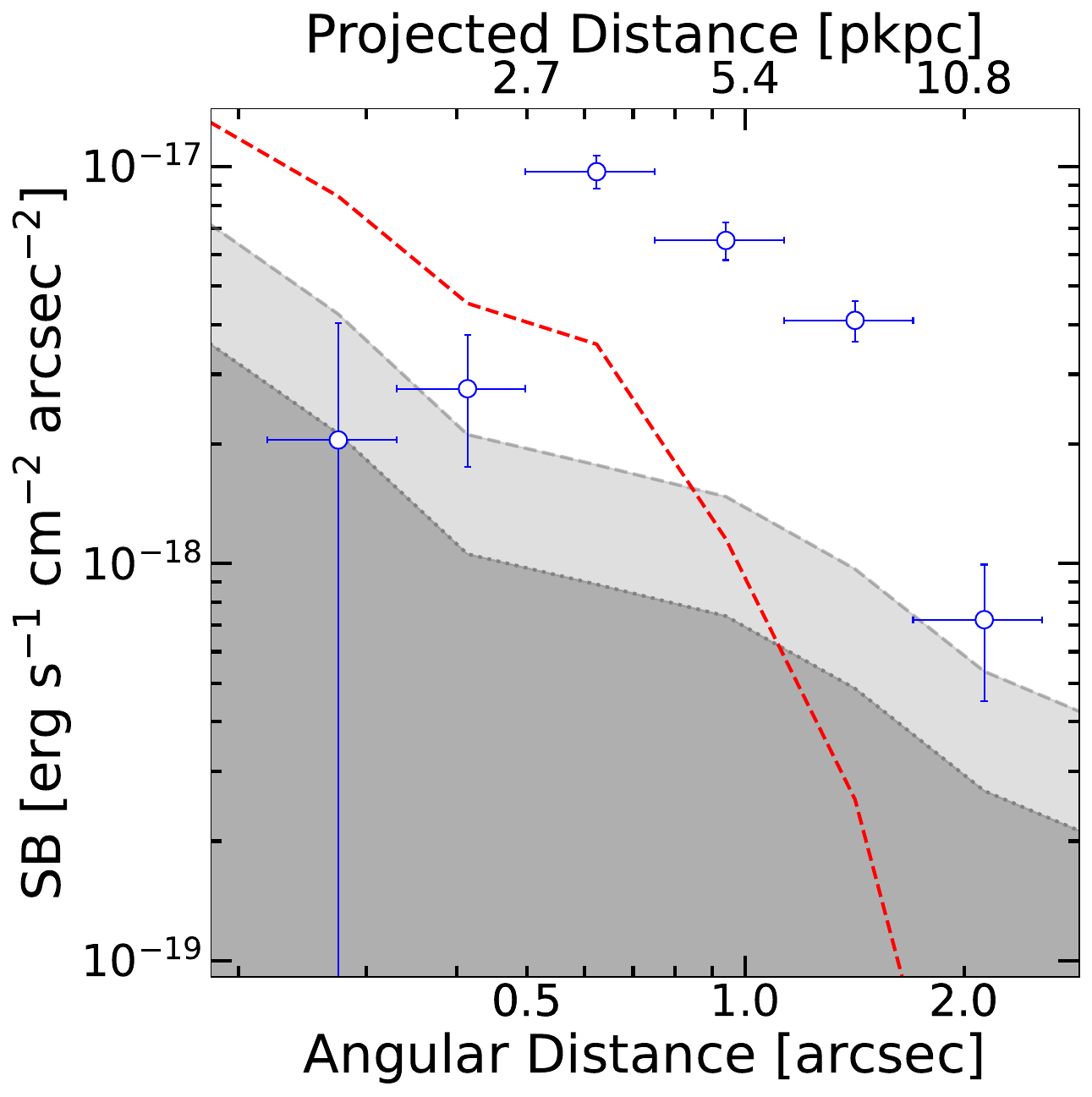}
    \caption{Radial surface brightness profile of the extended emission (blue points). The radial bins are logarithmically spaced, with the horizontal error bars spanning an entire bin. The shaded gray regions represent the $1\sigma$ (dark gray) and $2\sigma$ (light gray) noise levels of measured over $9750$\textendash$9780\;\mathrm{\AA}$. The PSF profile (FWHM$=1\farcs{0}$) is plotted as a red dashed line.}
    \label{fig:sb_profile}
\end{figure}

We derive the radial profile of the Ly$\alpha$ surface brightness by integrating our quasar-subtracted cube from $\lambda_{\mathrm{min}}$ to $\lambda_{\mathrm{max}}$, and calculating the surface brightness within circular annuli centered on the quasar. 
To compare the Ly$\alpha$ profile with the background noise level, we also calculate the surface brightness profile over the spectral region $9750$\textendash$9780\;\mathrm{\AA}$, where no extended emission is present. 
We selected this wavelength range because the quasar continuum is featureless, the spectral range is free of sky lines, and the $5\sigma$ observed surface brightness limit in empty sky regions is nearly identical to that in the Ly$\alpha$ spectral region ($5\times10^{-18}$ erg s$^{-1}$ cm$^{-2}$ arcsec$^{-2}$).
The radial surface brightness profile is displayed in Figure \ref{fig:sb_profile}.
We note that the extended emission is distributed asymmetrically about the quasar (e.g., black contours in Figure \ref{fig:j0910_multi_panel}), and that the profile shown in Figure \ref{fig:sb_profile} is circularly averaged.

\subsection{Kinematics of the Extended Ly$\alpha$} \label{sub:kin}

\begin{figure*}[ht]
    \centering
    \includegraphics[width=1\textwidth]{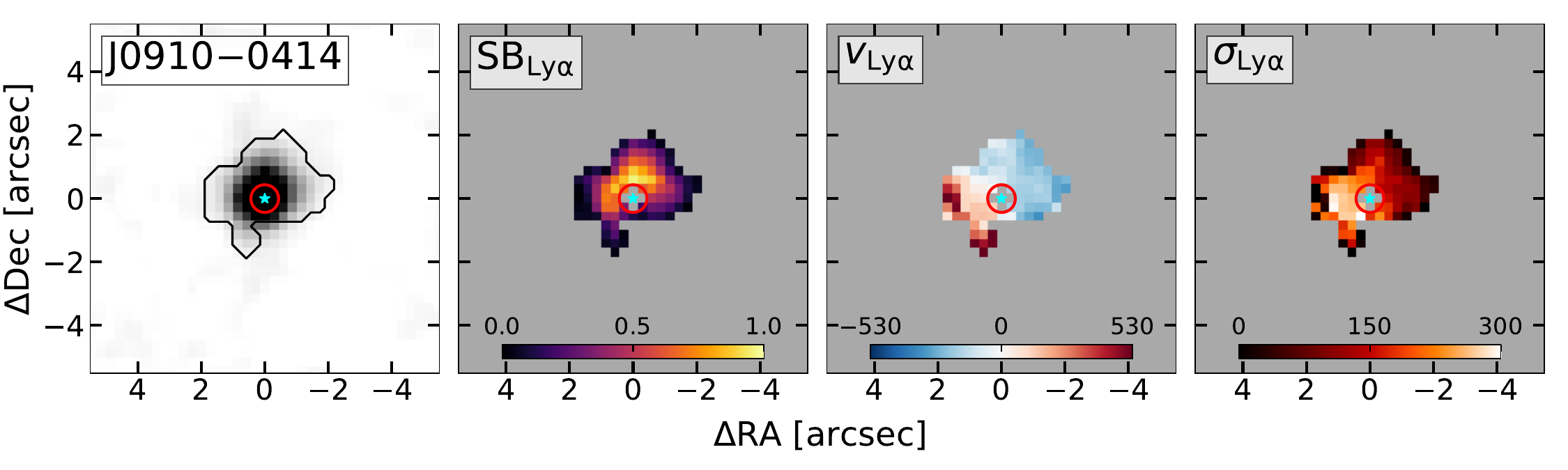}
    \caption{Moment maps of the extended Ly$\alpha$ around J0910$-$0414. The leftmost image was created by applying the 2D smoothing kernel from Section \ref{sub:kin} to our non-quasar-subtracted cube and collapsing over our Ly$\alpha$ wavelength range. The black contour outlines the 2D spatial mask, and is identical to the contour in Figure \ref{fig:j0910_multi_panel}. 
    The second panel is the observed Ly$\alpha$ surface brightness in units of $10^{-17}$ erg s$^{-1}$ cm$^{-2}$ arcsec$^{-2}$. 
    The third panel displays the velocity offset in km s$^{-1}$ relative to the centroid wavelength of the halo spectrum as described in Section \ref{sub:halo_spec}.
    The rightmost panel is the velocity dispersion in km s$^{-1}$. North is up and East is to the left.}
    \label{fig:lya_maps}
\end{figure*}

Characterizing the diffuse Ly$\alpha$ emission requires measurement of the spatially resolved kinematics. 
To facilitate comparison with results of the REQUIEM survey, we follow \cite{farina19} and convolve each wavelength slice with a 2D Gaussian kernel with $\sigma_{\mathrm{spat}} = 0\farcs{291}$ (1 pixel). 
This step is similar to the procedure applied to the S/N cube described in Section \ref{sec:detect}, but with no smoothing performed in the spectral direction. 
As a result, our maps are based on measurements performed on a spatially smoothed quasar-subtracted cube that does not contain artifacts (e.g., empty spaxels) from the cube creation step.
We calculate the velocity moments similarly to Section \ref{sub:halo_spec}, this time only including voxels contained in the 3D mask as described in Section \ref{sec:detect}. 
Velocities and dispersions were calculated relative to the Ly$\alpha$ centroid wavelength of the halo spectrum. 
Maps of the Ly$\alpha$ surface brightness, line velocity, and velocity dispersion are displayed in Figure \ref{fig:lya_maps}.

\begin{deluxetable}{lcc}
    \label{tab:halo}
    \tablecaption{J0910$-$0414 Ly$\alpha$ Halo Measurements}
    \tablecolumns{3}
    \tablehead{
    \colhead{Quantity} &
    \colhead{} &
    \colhead{Measured Value}
    }
    \startdata
    $\lambda_{\mathrm{min}}$, $\lambda_{\mathrm{max}}$  & & \phn$9277.1$  \AA, $9306.1$  \AA\\
    $d_{\mathrm{max}}$  & & $ 4\farcs{0}$/$21.3$ pkpc\phn\phn\phn\\
    $d_{\mathrm{r}}$   & & $ 2\farcs{1}$/$11.4$ pkpc\phn\phn\phn\\
    $A_{\mathrm{Ly\alpha}}^{\mathrm{halo}}$  & & \phn\phn\phn\phn\phn$7.6$ arcsec$^{2}$/$222.2$ pkpc$^{2}$\\
    $F_{\mathrm{Ly\alpha}}^{\mathrm{halo}}$  & & \phn\phn\phn\phn\phn\phn$(4.6\pm0.3)\times10^{-17}$ erg s$^{-1}$ cm$^{-2}$  \\
    $L_{\mathrm{Ly\alpha}}^{\mathrm{halo}}$ & & $(2.3\pm0.1)\times10^{43}$ erg s$^{-1}$\\
    $v_{\mathrm{Ly\alpha}}^{\mathrm{halo}}$ & & $203\pm 15$ km s$^{-1}$ \phn\\
    $\sigma_{\mathrm{Ly\alpha}}^{\mathrm{halo}}$ & & $242\pm9\phn$ km s$^{-1}$\phn  \\
    $\mathrm{FWHM}_{\mathrm{Ly\alpha}}^{\mathrm{halo}}$ &  & $474\pm74$ km s$^{-1}$\phn
    \enddata
    \tablecomments{Quantities derived from the 2D spatial mask and the corresponding halo spectrum. Conversions between units of arcsec and pkpc were performed using $1\arcsec=5.4\;\mathrm{pkpc}$.
    $d_{\mathrm{max}}$ corresponds to the largest separation between two pixels in the 2D spatial mask, and $d_{\mathrm{r}}$ is the largest distance between a pixel in the 2D spatial mask and the location of the quasar.
    $A_{\mathrm{Ly\alpha}}^{\mathrm{halo}}$ is the area of the 2D spatial mask. 
    $F_{\mathrm{Ly\alpha}}^{\mathrm{halo}}$, $v_{\mathrm{Ly\alpha}}^{\mathrm{halo}}$, and $\sigma_{\mathrm{Ly\alpha}}^{\mathrm{halo}}$ were derived from the velocity moments of the halo spectrum. The FWHM is calculated from the smoothed halo spectrum.
    All uncertainties reported in this table are statistical.
    }
\end{deluxetable}

\section{Results and Discussion} \label{sec:disc}

Our observations reveal extended Ly$\alpha$ surrounding J0910$-$0414 extending out to a radius of $\simeq11$ pkpc.
The low velocity and dispersion values relative to the halo's systemic redshift indicate quiescent kinematics consistent with expectations of gravitationally dominated motion.
Results from our moment maps are overall consistent with the properties of other halos detected in the Quasar Snapshot Observations with MUse: Search for Extended Ultraviolet eMission \citep[QSO MUSEUM, \added{$3.03\lesssim z\lesssim3.40$};][]{arrigoni19} and the REQUIEM (\added{$5.90 \lesssim z \lesssim 6.61$}) surveys.
\added{We note, however, that the effects of resonant scattering and attenuation by the intergalactic medium (IGM) at $z\sim6$ modify the Ly$\alpha$ profile \citep[e.g.,][]{dijkstra17}, complicating the interpretation of the halo gas kinematics.}
In the following subsections, we examine the properties of the extended halo, the spatially resolved Ly$\alpha$ kinematics, and possible mechanisms powering the observed halo.


\begin{figure*}[t]
    \centering
    \includegraphics[width=1\textwidth]{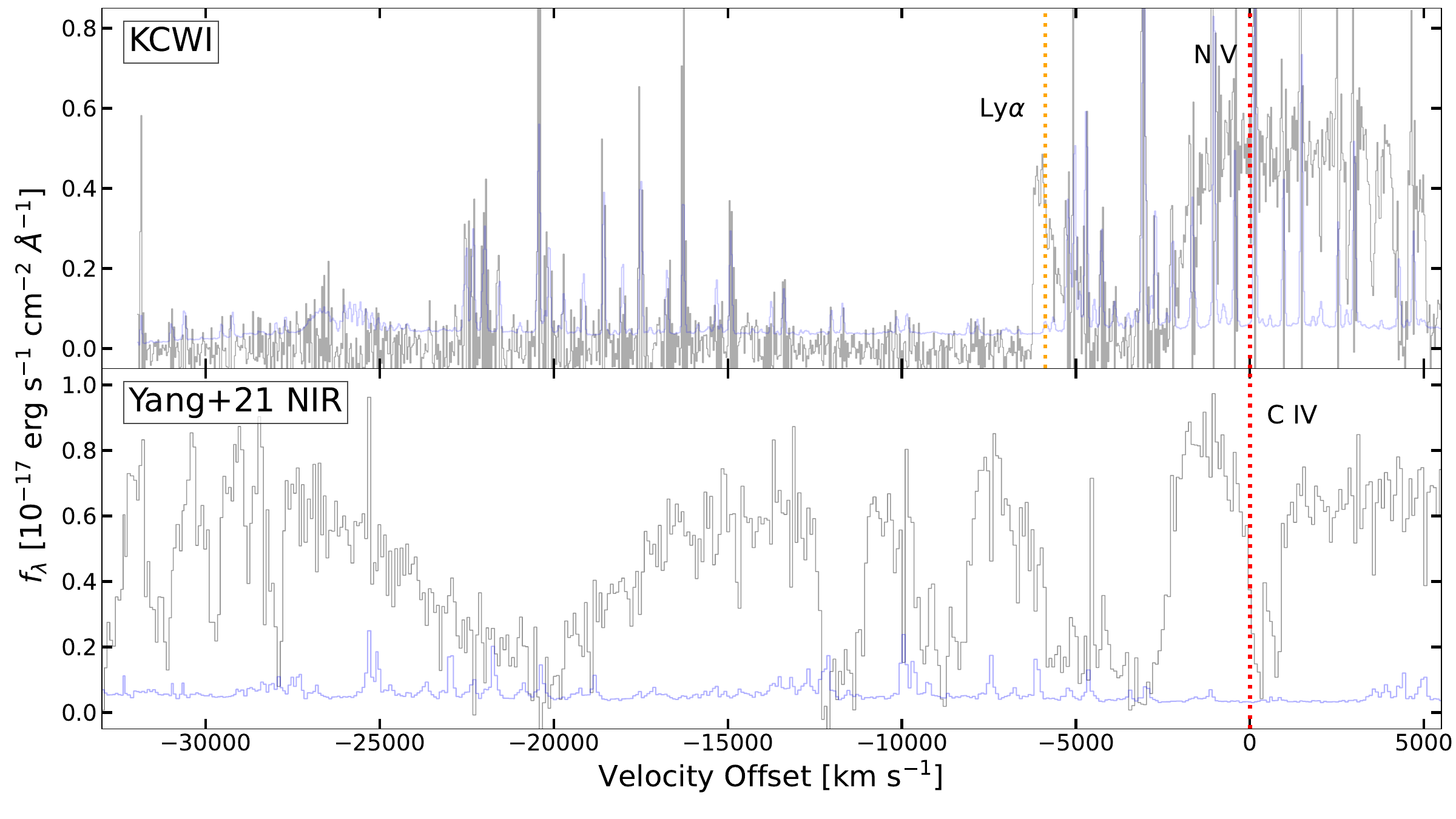}
\caption{
    Top panel: Total flux spectrum (black) and the $1\sigma$ error spectrum (blue) extracted from the 2D mask, plotted as a function of velocity offset from \ion{N}{5}. 
    The orange dotted line marks the velocity offset of the Ly$\alpha$ centroid with respect to \ion{N}{5}.
    Bottom panel: Total flux spectrum (black) and the $1\sigma$ error spectrum (blue) of the NIR spectrum from \cite{yang21}, plotted as a function from velocity offset from \ion{C}{4}.
    Values for the velocity range were chosen to cover the absoprtion line complex blueward of \ion{C}{4}.
    The red dotted line denotes the zero velocity offset, from \ion{N}{5} (top panel) or \ion{C}{4} (bottom panel). 
    }
    \label{fig:BAL}
\end{figure*}

\subsection{Spectrum of the extended halo}

Spectra of the total flux and halo emission are displayed in Figure \ref{fig:neb_spec} in gray and red, respectively. J0910$-$0414 was identified as a BAL quasar by \cite{wang19} based on its strong, blueshifted \ion{N}{5} absorption. 
Our spectrum of the total flux displays the same deep \ion{N}{5} absorption trough and the isolated Ly$\alpha$ peak.
From the total flux spectrum, we measure a Ly$\alpha$ integrated line flux of $F_{\mathrm{Ly \alpha}}^{\mathrm{total}}=(8.3\pm0.3)\times10^{-17} \; \mathrm{erg} \; \mathrm{s}^{-1} \; \mathrm{cm}^{-2}$.
Comparison with the line flux for the halo yields $ F_{\mathrm{Ly \alpha}}^{\mathrm{halo}} /  F_{\mathrm{Ly \alpha}}^{\mathrm{total}} = 0.55$, which means that $\gtrsim50\%$ of the integrated line flux can attributed to the extended halo. 
In other words, slightly over half of the Ly$\alpha$ flux detected within the 2D mask comes from the extended emission, rather from the quasar itself.
If the absorption trough blueward of \ion{N}{5} covers the Ly$\alpha$ wavelength range, then it is possible for most of the quasar flux in this spectral range to be obscured to some degree.
We show the total flux spectrum from KCWI and the NIR spectrum from \cite{yang21} in Figure \ref{fig:BAL}. 
The NIR spectrum displays multiple absorption troughs blueward of \ion{C}{4}, with the BAL complex extending out to $\Delta v_{\mathrm{C\;IV}} = -0.09c$ ($-27000$ km s$^{-1}$). 
Relative to the \ion{N}{5} line, Ly$\alpha$ is located at an offset of $\Delta v_{\mathrm{N\;V}} = -0.02c$ ($-6000$ km s$^{-1}$), corresponding in velocity to the blue end of the first \ion{C}{4} trough at $-5000$ km s$^{-1}$.
If we assume that the \ion{N}{5} and \ion{C}{4} BALs have similar velocity structure, then the Ly$\alpha$ 
\added{emission is} located where a BAL trough is starting to lift, thus allowing some nuclear emission to shine through.

\subsection{Size of the extended emission}
The halo surrounding J0910$-$0414 is distributed asymmetrically around the quasar with the majority of the emission located $\lesssim 2\farcs{0}$ ($\lesssim 11$ pkpc) toward the north. 
We measured \added{the maximum distance between two spaxels within the 2D spatial mask $d_{\mathrm{max}}$, and the distance from the quasar to the farthest spaxel within the 2D spatial mask $d_{\mathrm{r}}$, obtaining values of $d_{\mathrm{max}}=21.3$ pkpc and $d_{\mathrm{r}}=11.4$ pkpc, respectively.
Our measurements fall within the range of sizes reported for the detections in the REQUIEM sample.
}
\cite{wang24b} report effective sizes of $R_{\mathrm{[C \;II]}} = 1.48 \pm 0.19$ pkpc and $R_{\mathrm{cont}} = 0.62 \pm 0.05$ pkpc for the [\ion{C}{2}] and far infrared (FIR) continuum emission regions, respectively. 
These sizes are an order of magnitude smaller than $d_{\mathrm{r}}$ for the Ly$\alpha$ halo, but it should be noted that the measurements by \cite{wang24b} were obtained by fitting 2D Gaussians to maps of the [\ion{C}{2}] and continuum flux, while our measurement is based on the farthest spaxel from the quasar containing Ly$\alpha$ emission. 
If we instead use the largest radial bin of the radial surface brightness profile with a detection above the $2\sigma$ background noise level \citep[e.g., $d_{\mathrm{max}}^{\mathrm{Ly\alpha}}$ in][]{durovcikova25a}, we measure a nebula size of $d_{\mathrm{max}}^{\mathrm{Ly\alpha}}= 11.5 \pm2.3$ pkpc. 
Regardless, our measurements show that the size of the Ly$\alpha$ halo around J0910$-$0414 is approximately eight times the size of the [\ion{C}{2}] emission region, and 18 times the size of the FIR continuum emission region.
This large discrepancy in size indicates that Ly$\alpha$ traces the large gas reservoir surrounding the quasar, while the [\ion{C}{2}] and FIR continuum trace the gas and dust within the inner portion of the quasar host galaxy \citep{drake22}.

\subsection{Kinematics of the extended emission}
From the halo spectrum, the centroid of the extended Ly$\alpha$
emission yields a velocity shift of $v_{\mathrm{Ly\alpha}}^{\mathrm{halo}} = 203 \pm 15$ km s$^{-1}$ relative to the expected wavelength of the line at the [\ion{C}{2}] redshift. 
This low velocity offset relative to the [\ion{C}{2}]
\added{systemic} redshift signifies a close connection between the extended halo and the host galaxy of J0910$-$0414.
We measure nebular Ly$\alpha$ line widths of $\sigma_{\mathrm{Ly\alpha}}^{\mathrm{halo}} = 242 \pm 9$ km s$^{-1}$, and $\mathrm{FWHM}_{\mathrm{Ly\alpha}}^{\mathrm{halo}} = 474 \pm 74$ km s$^{-1}$, which are lower than the corresponding values for the vast majority of the REQUIEM sample, \added{but still consistent with expectations of gravitationally dominated motion (e.g., $\sigma_{\rm{Ly}\alpha} < 400$ km s$^{-1}$) within a dark matter halo}. 
\added{We reiterate that our kinematic measurements are affected by the effects of IGM attenuation, as well as resonant scattering.
The peak of Ly$\alpha$ at $z\sim6$ is typically absorbed by neutral hydrogen, which means that our measurement of $v_{\mathrm{Ly\alpha}}^{\mathrm{halo}}$ is likely redshifted relative to the intrinsic bulk velocity of the halo.
Resonant scattering likely contributes to the broadening of the line profile and can lead to larger line widths. 
While this strengthens the case that the halo gas kinematics are gravitationally dominated at lower redshifts \citep[e.g.,][]{arrigoni19}, attenuation of the Ly$\alpha$ profile means that these and subsequent results require confirmation from observations of a non-resonant line (e.g., H$\alpha$).}

\added{
Our measurement of $\mathrm{FWHM}_{\mathrm{Ly\alpha}}^{\mathrm{halo}}$ for J0910$-$0414 is notable when compared to the [\ion{C}{2}] line width \citep[$\mathrm{FWHM}_{\mathrm{[C\;II]}}=783\pm40$ km s$^{-1}$;][]{wang24b}. 
While FWHM differences between Ly$\alpha$ and [\ion{C}{2}] are expected (e.g., radiative transfer effects and attenuation of Ly$\alpha$, different regions within the overall gravitational potential, different gas phases),
halos detected in the REQUIEM survey consistently display $\mathrm{FWHM}_{\mathrm{Ly\alpha}}^{\mathrm{halo}} \gtrsim 2.0 \times \mathrm{FWHM}_{\mathrm{[C\;II]}}$ \citep[see Figure 5 of][]{farina19}. 
In contrast to this, J0910$-$0414 has $\mathrm{FWHM}_{\mathrm{Ly\alpha}}^{\mathrm{halo}} \simeq 0.6 \times \mathrm{FWHM}_{\mathrm{[C\;II]}}$.
This reversal could reflect dynamically hotter conditions for the [\ion{C}{2}]-emitting gas (e.g., from rapid star formation) in the compact host galaxy core relative to the Ly$\alpha$-emitting gas in the circumgalactic medium.
}

\subsection{Spatially Resolved Kinematics} \label{sub:spat_kin}

We display the spatially resolved kinematics of Ly$\alpha$, calculated relative to the halo's \added{systemic} redshift, in Figure \ref{fig:lya_maps}. 
Overall, the moment maps display ordered, quiescent kinematics. 
Velocity maps of extended Ly$\alpha$ tend to show no evidence of ordered motion (e.g., rotation, inflows, outflows), although the resonant nature of the Ly$\alpha$ emission may prevent the appearance of such kinematic signatures \citep[e.g.,][]{cantalupo05, farina19, drake22}. 
Notably, J1605$-$0112, another high-$z$ BAL quasar ($z=4.92$) having an extended Ly$\alpha$ halo, displays no clear pattern in its velocity field \citep[][]{ginolfi18}.
J0910$-$0414 displays a velocity gradient oriented from the southeast to the northwest, and spanning the range ($-330$ km s$^{-1}$, $+530$ km s$^{-1}$).
We note that the velocity dispersion map shows elevated values coincident with this velocity gradient, so it is possible that the observed dispersions result in part from unresolved velocity gradients. 

\added{Rotation-like velocity shears, while rare among extended Ly$\alpha$ halos, have been reported in other systems, including J1020$+$1040 from QSO MUSEUM, and P231$-$20 from the REQUIEM sample \citep[][]{arrigoni18,farina19}, both of which also reside in overdense regions \citep{arrigoni18, meyer22, wang24a}. 
However, while such velocity gradients can arise from a variety of physical scenarios \citep[e.g.,][]{arrigoni18}, the limited extent, spatial resolution, and S/N of the halo presented here do not allow us to identify its physical origin.
Furthermore, the effects of resonant scattering and IGM attenuation at $z\sim6$ complicate the interpretation of the Ly$\alpha$ kinematics.
A clear interpretation of the halo kinematics requires deep, high-resolution observations of a non-resonant emission line.
}

ALMA observations of the host galaxy [\ion{C}{2}] emission of J0910$-$0414 by \cite{wang24b} reveal a rotation-like velocity gradient oriented North-South, spanning the range ($-100 \; \mathrm{km} \; \mathrm{s}^{-1}$, $+100$ km s$^{-1}$), as well as high velocity dispersions ($ 200 \; \mathrm{km} \; \mathrm{s}^{-1} \lesssim \sigma_{\mathrm{ [C\;II] }} \lesssim 300 \; \mathrm{km} \; \mathrm{s}^{-1} $).
The contrast between the orientations of the velocity gradients as well as the line width measurements demonstrate that the Ly$\alpha$-emitting gas in the surrounding reservoir and the [\ion{C}{2}]-emitting gas in the compact core of the host galaxy are kinematically decoupled, consistent with findings of other quasars at $z\sim6$ \citep{drake22}.

Previous observations and simulations indicate that extended Ly$\alpha$ halos around quasars may be powered by Ly$\alpha$ fluorescence (recombination emission of gas photoionized by the quasar), collisional excitation, resonant scattering of Ly$\alpha$, or a combination of these mechanisms \citep[e.g.,][]{hennawi_prochaska_13,cantalupo17, costa22, hoshi25}.
Ly$\alpha$ fluorescence is believed to be the dominant mechanism powering the halos \citep[e.g.,][]{durovcikova25b}, and our measured halo luminosity for J0910$-$0414 is within the typical range of values at $z\sim6$ ($L_{\mathrm{Ly \alpha}}\lesssim 10^{44}$ erg s$^{-1}$) expected for fluorescence from optically thin clouds in the gas reservoir \citep{farina19}.
\added{Furthermore, the halo of J0910$-$0414 is consistent with the correlation between the Ly$\alpha$ halo luminosity and quasar ionizing luminosity reported by \cite{hoshi25}.
These findings suggest that similar to other halos at around quasars at $z\sim6$, multiple mechanisms may be powering the halo around J0910$-$0414.}
Given the SFR estimates by \cite{wang24b}, it is also possible that the rapid star formation in J0910$-$0414's host galaxy contributes to the observed halo.

\section{Conclusions and Future Work} \label{sec:end}
From our ongoing KCWI program to search for  Ly$\alpha$ halos around $z>6.5$ quasars, we have detected significant extended Ly$\alpha$ emission out to $\simeq11$ pkpc around the BAL quasar J0910$-$0414. 

This result demonstrates the potential of KCWI's red channel for detecting the low surface brightness emission at redshifts that were previously inaccessible with ground-based integral-field instruments.
Future papers will include analysis for the rest of our sample,
extending our search range to $z=7.1$.
The capabilities of KCWI's red channel provide a unique opportunity to probe the fuel supply powering these early quasars, and to push the frontier of extended Ly$\alpha$ searches deeper into the epoch of reionization.

\begin{acknowledgments}

\added{We would like to thank the anonymous referee for their comments, which have improved the quality of this manuscript.}

RR would like to thank the \texttt{PypeIt} team for their efforts to adapt the software to work on data from KCWI's red channel.

FW acknowledges support from NSF award AST-2513040.

EPF is supported by the international Gemini Observatory, a program of NSF NOIRLab, which is managed by the Association of Universities for Research in Astronomy (AURA) under a cooperative agreement with the U.S. National Science Foundation, on behalf of the Gemini partnership of Argentina, Brazil, Canada, Chile, the Republic of Korea, and the United States of America.

The data presented herein were obtained at Keck Observatory, which is a private 501(c)3 non-profit organization operated as a scientific partnership among the California Institute of Technology, the University of California, and the National Aeronautics and Space Administration. The Observatory was made possible by the generous financial support of the W. M. Keck Foundation.

This work was supported by a NASA Keck PI Data Award, administered by the NASA Exoplanet Science Institute. Data presented herein were obtained at the W. M. Keck Observatory from telescope time allocated to the National Aeronautics and Space Administration through the agency's scientific partnership with the California Institute of Technology and the University of California. The Observatory was made possible by the generous financial support of the W. M. Keck Foundation.

The authors wish to recognize and acknowledge the very significant cultural role and reverence that the summit of Maunakea has always had within the Native Hawaiian community. We are most fortunate to have the opportunity to conduct observations from this mountain.

This research has made use of the Keck Observatory Archive (KOA), which is operated by the W. M. Keck Observatory and the NASA Exoplanet Science Institute (NExScI), under contract with the National Aeronautics and Space Administration.

\end{acknowledgments}

%
\facility{Keck:II (KCWI)}

\software{astropy \citep{astropy:2013, astropy:2018, astropy:2022},
Matplotlib \citep{Hunter:2007},
NumPy \citep{harris2020array},
SciPy \citep{2020SciPy-NMeth},
PypeIt \citep{pypeit_release, pypeit_joss},
dust\_extinction \citep{Gordon2024}
}


\bibliography{j0910}{}
\bibliographystyle{aasjournalv7}

\end{document}